\documentclass[aip,jmp,preprint,onecolumn,showpacs,amsmath,amssymb]{revtex4-1}
\usepackage{graphicx}
\usepackage{bm}
\usepackage[colorlinks=true,linkcolor=black, citecolor=blue, urlcolor=blue,letterpaper,bookmarks=true,bookmarksopen=false,bookmarksnumbered=false,hypertexnames=true]{hyperref}

\newcommand{\pd}[1]{\partial_{#1}}
\newcommand{\bu}{\bm{u}}
\newcommand{\ov}[1]{\overline{#1}}
\newcommand{\wRa}{\widetilde{Ra}}
\newcommand{\eps}{\epsilon}
\newcommand{\ddz}[1]{\frac{\text{d}#1}{\text{d}z}}
\newcommand{\ddzt}[1]{\frac{\text{d}^2#1}{\text{d}z^2}}

\begin{document}
\preprint{Nonlinearity} 
\title{Asymptotic Behavior of Heat Transport for a Class of Exact Solutions in Rotating Rayleigh-B\'{e}nard Convection} 
\author{Ian Grooms}\email{grooms@cims.nyu.edu} \affiliation{Courant Institute of Mathematical Sciences, New York University, New York, New York 10012-1185 USA}

\date{\today} 

\begin{abstract} 
The non-hydrostatic, quasigeostrophic approximation for rapidly rotating Rayleigh-B\'enard convection admits a class of exact `single mode' solutions.
These solutions correspond to steady laminar convection with a separable structure consisting of a horizontal planform characterized by a single wavenumber multiplied by a vertical amplitude profile, with the latter given as the solution of a nonlinear boundary value problem.
The heat transport associated with these solutions is studied in the regime of strong thermal forcing (large reduced Rayleigh number $\wRa$).
It is shown that the Nusselt number $Nu$, a nondimensional measure of the efficiency of heat transport by convection, for this class of solutions is bounded below by $Nu\gtrsim \wRa^{3/2}$, independent of the Prandtl number, in the limit of large reduced Rayleigh number.
Matching upper bounds include only logarithmic corrections, showing the accuracy of the estimate.
Numerical solutions of the nonlinear boundary value problem for the vertical structure are consistent with the analytical bounds.
\end{abstract}
\maketitle 

\section{Introduction}
Thermal fluid convection influenced by rotation occurs in planetary and stellar atmospheres and in the Earth's molten core.
Rayleigh-B\'enard convection is an idealized setting for the exploration of convection; it consists of a layer of fluid between cold top and hot bottom boundaries held at constant temperature.
The efficiency of convection is measured by the Nusselt number $Nu$, which is the ratio of the total heat transport to the transport that would be affected by conduction alone. 
In the presence of rotation about a vertical axis, the dynamics are governed by three nondimensional numbers, the Rayleigh, Ekman, and Prandtl numbers
\[Ra=\frac{g\alpha_T(\Delta\!T)H^3}{\nu\kappa},\quad E=\frac{\nu}{2\Omega H^2},\quad\sigma=\frac{\nu}{\kappa}.\]
The kinematic viscosity is $\nu$, $\kappa$ is the thermal diffusivity, $g$ is the rate of gravitational acceleration, $H$ is the distance between the top and bottom boundaries, $\Omega$ is the system rotation rate, $\alpha_T$ is the thermal expansion coefficient, and $\Delta\!T$ is the magnitude of the temperature difference between the boundaries.
The Taylor number $Ta=E^{-2}$ is sometimes used in place of the Ekman number.

System rotation can have a profound impact on the fluid dynamics, e.g.~by shutting off convection for sufficiently fast rotation at fixed thermal forcing.
The critical Rayleigh number for the onset of convection increases as $Ra\sim E^{-4/3}$, and the wavenumber of the most unstable mode increases as $k\sim E^{-1/3}$; the linear stability properties of rotating Rayleigh-B\'enard convection are summarized in Refs.~\onlinecite{Chandra53} and \onlinecite{ChandraBook}.

Inspired by the scaling of the linear instability, reduced non-hydrostatic quasigeostrophic equations (NHQGE) for thermal convection, equations (1.a-d) below, were derived in Ref.~\onlinecite{JKW98}.
The NHQGE are derived in the limit of small Ekman numbers with the Rayleigh number scaled to $\wRa = Ra E^{4/3}$; the horizontal length scales are also scaled with the Ekman number as $L=E^{1/3}H$ where $H$ is the depth of the layer.
The NHQGE have also been generalized to situations where the axis of rotation does not align with gravity.\cite{JKMW06}

In the context of Rayleigh-B\'enard convection the NHQGE admit exact `single mode' solutions that have provided a useful point of comparison for simulations of the turbulent dynamics.\cite{JKW98,SJKW06,GJKW10,JRGK12}
The solutions consist of a separable ansatz (equation (\ref{eqn:Ansatz})) where all fields share the same horizontal structure multiplied by vertical amplitude functions that are given as the solution of a nonlinear two-point boundary value problem.
Examples of the horizontal structure include repeating patterns of convection rolls, squares, or hexagons.\cite{JK99}
This ansatz has a long history for the unreduced equations, including for example Refs.~\onlinecite{Veronis59,GST75,TSG77,BZ94}.
In Ref.~\onlinecite{BZ94} the ansatz was also shown to produce accurate approximate solutions of the unreduced Boussinesq equations at large Rayleigh and small Ekman numbers even for strongly nonlinear convection, well beyond the usual weakly-nonlinear theory.

The asymptotic behavior of these solutions at large Rayleigh numbers is studied here.
Upper and lower bounds on the Nusselt number are derived, showing that the asymptotic behavior of the Nusselt number for this class of solutions is at least as large as $\wRa^{3/2}$, but must be smaller than $\wRa^{3/2+\eps}$ for any $\eps>0$.
For a fixed wavenumber $k$ (independent of $\wRa$) the Nusselt number is asymptotically bounded by $\wRa\lesssim Nu\lesssim \wRa\ln(\wRa Nu)$.
(Note that constant pre-factors in asymptotic expressions are generally omitted throughout the paper for clarity.)
Faster growth is achieved by allowing the wavenumber $k$ to grow with $\wRa$.
Preliminaries, lower, and upper bounds are presented in the following sections, followed by some numerical solutions, and finally by further discussion of the results in the last section.

\section{Preliminaries}
The non-hydrostatic quasigeostrophic equations for rotating Rayleigh B\'enard convection are\cite{JKW98,SJKW06}
\begin{align} 
\pd{t}w+J[\psi,w]+\pd{z}\psi &=\frac{\wRa}{\sigma}\theta + \nabla_h^2w \tag{1.a}\\
\pd{t}\zeta + J[\psi,\zeta]-\pd{z}w &=\nabla_h^2\zeta\tag{1.b}\\
\pd{t}\theta+J[\psi,\theta]+w\pd{z}\ov{T}&=\frac{1}{\sigma}\nabla_h^2\theta\tag{1.c}\\
\pd{\tau}\ov{T}+\pd{z}\left(\ov{w\theta}\right)&=\frac{1}{\sigma}\pd{z}^2\ov{T}.\tag{1.d}
\end{align}
Boundary conditions at $z=0$ and $1$ are $w=\theta=\pd{z}\psi=0$, and $\ov{T}(0)=1$, $\ov{T}(1)=0$.
The vertical velocity is $w$, $\zeta$ is the vertical component of vorticity and is related to the geostrophic streamfunction $\psi$ for the horizontal velocities by $\nabla_h^2\psi=\zeta$.
Advection is purely horizontal and is written using the Jacobian operator $J[\psi,(\cdot)]=\bu\cdot\nabla(\cdot)$ where $u=-\pd{y}\psi$ and $v=\pd{x}\psi$.
The temperature is split into a horizontal mean $\ov{T}$ and a deviation $\theta$ of order $E^{1/3}$, and the mean temperature evolves on a slower time coordinate $\tau$. 
The overbar $\ov{(\cdot)}$ denotes an average over the horizontal coordinates and the fast time $t$.
The system can be written with only one time coordinate by replacing $\pd{\tau}=E^{-2/3}\pd{t}$, but this is not necessary for the following.


Equations for infinite Prandlt number convection\cite{SJKW06} may be derived by rescaling time such that $\pd{t}\to\sigma^{-1}\pd{t},\;\pd{\tau}\to\sigma^{-1}\pd{\tau}$, rescaling the velocities $\psi\to\sigma^{-1}\psi$, $w\to\sigma^{-1}w$ and then taking $\sigma\to\infty$.
The result is
\begin{align*}
\pd{z}\psi &=\wRa\theta + \nabla_h^2w \\
-\pd{z}w &=\nabla_h^2\omega\\
\pd{t}\theta+J[\psi,\theta]+w\pd{z}\ov{T}&=\nabla_h^2\theta\\
\pd{\tau}\ov{T}+\pd{z}\left(\ov{w\theta}\right)&=\pd{z}^2\ov{T}.
\end{align*}
These have the same form as the $\sigma=1$ equations without the inertial terms.
The same result is reached by first taking the infinite Prandtl number limit of the Boussinesq equations and then taking the non-hydrostatic quasigeostrophic limit.\\

There are exact steady solutions, at any Rayleigh and Prandtl number, that have the form
\setcounter{equation}{1}
\begin{equation}\label{eqn:Ansatz}
\left(\begin{array}{c}w\\\psi\\\theta\end{array}\right) = 
\left(\begin{array}{c}W(z)\\\Psi(z)\\\Theta(z)\end{array}\right)h(x,y)
\end{equation}
where $h(x,y)$ is called the `planform' and satisfies $\nabla_h^2h=-k^2h$ (with $k>0$) and $\ov{h^2}=1$.
The nonlinearities vanish for this ansatz because $J[h,h]=0$.
Any sum of Fourier modes with wavenumbers of the same magnitude is a planform.
Solutions of this type are discussed in Refs.~\onlinecite{JKW98,JK99,SJKW06,JK07,GJKW10,JRGK12}.
The vertical structure satisfies
\begin{equation}
\begin{split}
\ddz{\Psi} =\frac{\wRa}{\sigma}\Theta-k^2W,\qquad-\ddz{W} =k^4\Psi,\\
W\ddz{\ov{T}}=-\frac{k^2}{\sigma}\Theta,\qquad\pd{z}\left(W\Theta\right)=\frac{1}{\sigma}\ddzt{\ov{T}}.
\end{split}
\end{equation}
The dependence on Prandtl number $\sigma$ can be removed by the rescaling $W\to W/\sigma$ and $\Psi\to\Psi/\sigma$; the resulting equations also apply to the infinite Prandtl number model.
For the remainder of the discussion the notation is simplified by setting $\sigma=1$ without loss of generality.

The vertical structure equations may be condensed into the following nonlinear boundary value problem
\begin{gather}
\left[\ddzt{}+k^2\left(\frac{\wRa Nu}{1+k^{-2}W^2}-k^4\right)\right]W=0,\label{eqn:SM}\\
Nu = \left(\int_0^1\frac{\text{d}z}{1+k^{-2}W^2}\right)^{-1}.\label{eqn:Nu}
\end{gather}
Note that the mean temperature profile can be recovered by integrating 
\begin{equation}
\ddz{\ov{T}}= -\frac{Nu}{1+k^{-2}W^2}.\label{eqn:DT}
\end{equation}
These are exactly the same vertical structure equations derived in Ref.~\onlinecite{BZ94} for approximate solutions of the rotating Boussinesq equations at large Rayleigh and small Ekman numbers.
Numerical solutions of these equations for various $k$ and $\wRa$ can be found in a variety of references,\cite{BZ94,SJKW06,GJKW10,JRGK12} and in section \ref{sec:Num} below.

In the following it will be convenient to define
\begin{equation*}
\gamma = \wRa Nu.
\end{equation*}

Multiplying (\ref{eqn:SM}) by $W'(z)$ results in an exact differential, which integrates as follows
\begin{equation}
\left(\ddz{W}\right)^2 + k^4\gamma\ln(1 + k^{-2}W^2) - k^6W^2 = c^2.
\end{equation}
Note that at $W(0)=0$ so $c^2=W'(0)^2$ (also at $z=1$).

There are two solution branches, positive and negative.
Solutions must ascend one branch until the vertical velocity reaches a maximum $W=W_m$ where $W'=0$, and then switch branches to return back to $W=0$.
This switching can happen several times over the interval $z\in[0,1]$; such solutions are analogous to the infinitesimal solutions near the onset of convection (d$\ov{T}/$d$z=-1$) which have the form $W(z)\sim\sin(n\pi z)$.
Using the new notation $W_m$ allows the vertical structure equation to be written as 
\begin{equation}
\left(\ddz{W}\right)^2 + f(W) = f(W_m)\label{eqn:Squared}
\end{equation}
where 
\begin{equation}
f(W) = k^4\gamma\ln(1 + k^{-2}W^2) - k^6W^2.\label{eqn:fW}
\end{equation}
Note that $f(W)$ attains a maximum at $W=W_*$ which satisfies
\begin{equation*}
W_*^2 = \frac{\gamma}{k^2}-k^2.
\end{equation*}

Thus far the equations admit the trivial solution $W=0$, $Nu=1$ at any value of $\gamma$ and $k$.
The trivial solution can be ruled out by requiring $W$ to reach a nonzero maximum $W_m>0$.
In particular, a solution that ascends from the boundary to reach a peak at mid layer must have
\begin{equation}
\int_0^{W_m} \frac{\text{d}W}{\left[f(W_m)-f(W)\right]^{1/2}} = \frac{1}{2}.\label{eqn:ZInt}
\end{equation}
This integral will not converge for $W_m\ge W_*$, so $W_*$ is an upper bound for $W_m$.
Extension to solutions that oscillate across the layer is straightforward, replacing the right hand side by $1/(2n)$ where $n$ is the number of oscillations.

Note that the equation for the Nusselt number (\ref{eqn:Nu}) can be written as an integral against d$W$ as follows
\begin{align}\notag
Nu^{-1} &= \int_0^{1/2} \frac{\text{d}z}{1 + k^{-2}W^2}+\int_{1/2}^1 \frac{\text{d}z}{1 + k^{-2}W^2}\\\notag
&=\int_0^{W_m} \frac{\text{d}W}{(1 + k^{-2}W^2)\left[f(W_m)-f(W)\right]^{1/2}}
-\int_{W_m}^0 \frac{\text{d}W}{(1 + k^{-2}W^2)\left[f(W_m)-f(W)\right]^{1/2}}\\
&=2\int_0^{W_m} \frac{\text{d}W}{(1 + k^{-2}W^2)\left[f(W_m)-f(W)\right]^{1/2}}.\label{eqn:NuInt}
\end{align}
The behavior of these solutions at large $\wRa$ is investigated in the next section.
For a solution that oscillates $n$ times between the boundaries the above equation is simply multiplied by $n$, implying that the Nusselt number for such solutions is smaller than for solutions with a single rise and fall between the boundaries.

Equations (\ref{eqn:ZInt}) and (\ref{eqn:NuInt}) do not guarantee the existence of nontrivial single mode solutions; rather, they describe properties of such solutions if they exist.

\section{Asymptotics}
\subsection{Bounds on $k$\label{sec:3A}}
Note that (\ref{eqn:Nu}) implies
\begin{equation}
Nu \le 1+\left(\frac{W_m}{k}\right)^2.
\end{equation}
Since $W_m^2$ is bounded above by $W_*^2$ (as noted below equation (\ref{eqn:ZInt}) above), this further implies
\begin{equation}
Nu < 1+\left(\frac{W_*}{k}\right)^2 = \frac{\wRa Nu}{k^4},
\end{equation}
and finally
\begin{equation}
k<\wRa^{1/4},\label{eqn:kUpper}
\end{equation}
i.e.~there are no nontrivial solutions for $k\ge\wRa^{1/4}$.\\

Next consider the behavior at small $k$.
Use equation (\ref{eqn:DT}) to rewrite equation (\ref{eqn:SM}) as
\begin{equation}
\left[\ddzt{}-k^2\left(\wRa\ddz{\ov{T}}+k^4\right)\right]W=0,
\end{equation}
then multiply by $W$ and integrate to arrive at
\begin{equation}
\left\|\ddz{W}\right\|_2^2 + k^6\|W\|_2^2 +k^2\wRa\int_0^1\ddz{\ov{T}}W^2\text{d}z=0.
\end{equation}
(Here and throughout $\|\cdot\|_2$ denotes the $L^2$ norm for functions on $z\in(0,1)$.)
An integration by parts yields
\begin{equation}
\left\|\ddz{W}\right\|_2^2 + k^6\|W\|_2^2 -2k^2\wRa\int_0^1\ov{T}W\ddz{W}\text{d}z=0.\label{eqn:16}
\end{equation}
The amplitude of the last term can be bounded by noting that $0\le\ov{T}\le1$, which is guaranteed by the negativity of equation (\ref{eqn:DT}) together with the boundary conditions on $\ov{T}$, and by using a version of Young's inequality ($2ab\le a^2+b^2$):
\begin{align*}
2\int_0^1 \ov{T}W\ddz{W}\text{d}z&\le2\int_0^1 |W|\left|\ddz{W}\right|\text{d}z\le\pi \|W\|^2+\pi^{-1}\left\|\ddz{W}\right\|^2.
\end{align*}
Together with equation (\ref{eqn:16}) this implies
\begin{equation}
\left(1-k^2\wRa\pi^{-1}\right)\left\|\ddz{W}\right\|_2^2 + k^2\left(k^4-\wRa\pi\right)\|W\|_2^2\le0.
\end{equation}
This inequality must be satisfied by any nontrivial solution of the single mode equations.
Consider the case of large horizontal scales, specifically where $k\ll\wRa^{-1/2}$; for these wavenumbers $1-k^2\wRa\pi^{-1}\ge0$ and application of the Poincar\'e inequality $\|\text{d}W/\text{d}z\|_2^2\ge\pi^2\|W\|_2^2$ to the above yields
\begin{equation}
\left(\left(\pi^2-k^2\wRa\pi\right)+ k^2\left(k^4-\wRa\pi\right)\right)\|W\|_2^2\le0.
\end{equation}
A nontrivial solution must therefore have
\begin{equation}
2\pi\wRa \ge\frac{\pi^2}{k^2}+k^4,
\end{equation}
but this condition cannot be met for $k\ll\wRa^{-1/2}$, therefore there can be no nontrivial solutions for $k\ll\wRa^{-1/2}$.

This analysis agrees qualitatively with the marginal stability curve for the onset of steady (as opposed to oscillatory, see e.g.~Ref.~\onlinecite{JK99}) convection; the linear stability calculation can be found in, e.g.~Ref.~\onlinecite{SJKW06}.
Specifically, the conduction solution $\ov{T}=1-z$ is stable to infinitesimal normal-mode perturbations with wavenumber $k$ provided that
\begin{equation}
\wRa < \frac{\pi^2}{k^2} + k^4.\label{eqn:LinStab}
\end{equation}
For large $\wRa$ there is a finite interval where single mode solutions can exist; for large $\wRa$ the interval is asymptotically contained in $k\in(\wRa^{-1/2},\wRa^{1/4})$.

\subsection{Bounds on $W$}
Make the following change of variable: $W = k v$.
Then
\[f(W) = f(kv) = k^4\gamma\ln(1 + v^2) - k^8v^2\]
and the condition that $W$ reaches its maximum at mid layer, equation (\ref{eqn:ZInt}), becomes
\begin{equation}
\int_0^{v_m} \frac{\text{d}v}{\left[\ln((1+v_m^2)/(1+v^2))-k^4\gamma^{-1}(v_m^2-v^2)\right]^{1/2}}= \frac{k\gamma^{1/2}}{2}.\label{eqn:Half}
\end{equation}
The condition $W_m<W_*$ implies
\begin{equation}
\gamma > k^{4}(1+v_m^2)\label{eqn:vmAbove}
\end{equation}
which allows the integral to converge.

First note that $v_m$ must go to infinity as $\gamma\to\infty$, which can be proven by a reductio argument as follows.
Suppose that $v_m$ remains bounded but $\gamma\to\infty$.
Furthermore, consider $k\sim\gamma^\alpha$ for $-1/2<\alpha<1/4$, compatible with the foregoing bounds on $k$.
Then the RHS of equation (\ref{eqn:Half}) grows to infinity, while the left hand side remains bounded.
Thus, $v_m$ cannot be bounded above as $\gamma\to\infty$.
Note that this does not guarantee the existence of solutions; rather, if nontrivial solutions exist for $k\sim\gamma^\alpha$ with $-1/2<\alpha<1/4$ then they must have $v_m\to\infty$ as $\gamma\to\infty$.

Now the integral (\ref{eqn:Half}) can be used to develop a lower bound for $v_m$.
The radicand of the denominator can be bounded as follows
\begin{equation}
\ln((1+v_m^2)/(1+v^2))-k^4\gamma^{-1}(v_m^2-v^2)\ge 2v_m((1+v_m^2)^{-1}-k^4(\gamma^{-1})(v_m-v)
\end{equation}
which is valid on the interval $v\in[0,v_m]$ for $v_m$ above a threshold of approximately $v_m>1.98$.
This and all such bounds used throughout this section can be trivially proven by showing that the sign of the error is correct (either positive or negative as necessary) over the interval $v\in[0,v_m]$.
It suffices to check the sign of the error (or of the first nonzero derivative, if the sign is zero) at the endpoints of the interval and at any critical points that lie in the interval.

The resulting bound on the integral is
\begin{align}
\frac{k\gamma^{1/2}}{2}&\le \int_0^{v_m} \frac{\text{d}v}{\left[2v_m((1+v_m^2)^{-1}-k^4\gamma^{-1}(v_m-v)\right]^{1/2}}=\left(\frac{2(1+v_m^2)}{1-(1+v_m^2)k^4\gamma^{-1}}\right)^{1/2}.
\end{align}
This implies
\begin{equation}
\frac{\gamma}{k^4+8k^{-2}}\le 1+v_m^2.\label{eqn:vmBelow}
\end{equation}
Consider the case where $k=K\gamma^{\alpha}$, with $-1/2<\alpha<1/4$, i.e.
\begin{equation*}
\frac{\gamma}{(K\gamma)^{4\alpha}+8(K\gamma)^{-2\alpha}}\le 1+v_m^2.
\end{equation*} 
For $\alpha=0$ this bound asymptotically pinches the upper bound, giving $v_m^2\sim \gamma$, but for other $\alpha$ the precise rate of increase of $v_m$ with $\gamma$ is not known.
It is noted in Ref.~\onlinecite{SJKW06} that equation (\ref{eqn:DT}) implies that, for fixed $k$, the mean temperature gradient at mid layer scales as $\wRa^{-1}$, i.e.~an isothermal interior develops at large Rayleigh numbers.
The above bound only substantiates this result at fixed $k$.

\subsection{Lower Bounds on $Nu$}
Under the change of variable $W=kv$ equation (\ref{eqn:NuInt}) for the Nusselt number becomes
\begin{equation}Nu^{-1} =\frac{2}{k\gamma^{1/2}}\int_0^{v_m}\frac{\text{d}v}{(1+v^2)\left[\ln((1+v_m^2)/(1+v^2))-k^4\gamma^{-1}(v_m^2-v^2)\right]^{1/2}}.\label{eqn:NuV}
\end{equation}
This can be bounded using the following lower bound to the radicand of the denominator of (\ref{eqn:NuV}), which is valid for sufficiently large $v_m$
\begin{equation}
\ln\left(\frac{1+v_m^2}{1+v^2}\right)-\frac{k^4(v_m^2-v^2)}{\gamma}
\ge \frac{2v_m}{1+v_m^2}\Delta(v_m-v) 
+ \frac{\left(v_m^2-1\right) (v-v_m)^2}{\left(1+v_m^2\right)^2}
\end{equation}
where
\begin{equation*}
 \Delta = 1-\frac{k^4(1+v_m^2)}{\gamma}.
\end{equation*}
Note that the bound (\ref{eqn:vmAbove}) implies that $1>\Delta>0$.
The integral that results from inserting this approximation into (\ref{eqn:NuV}) can be evaluated exactly, giving
\begin{equation}
Nu^{-1}\le \frac{2 \left( v_m^2+1\right)}{\sqrt{\gamma } k} \left(\frac{\tan ^{-1}\left(\frac{\sqrt{ v_m-i} \sqrt{ v_m^2+2 \Delta  ( v_m+i)  v_m-1}}{\sqrt{-( v_m-i) \left(2 \Delta  \left( v_m^2+1\right)+ v_m^2-1\right)}}\right)}{( v_m-i) \sqrt{ v_m^2+2 \Delta  ( v_m+i)  v_m-1}}+\frac{\tan ^{-1}\left(\frac{\sqrt{ v_m+i} \sqrt{ v_m^2+2 \Delta  ( v_m-i)  v_m-1}}{\sqrt{-( v_m+i) \left(2 \Delta  \left( v_m^2+1\right)+ v_m^2-1\right)}}\right)}{( v_m+i) \sqrt{ v_m^2+2 \Delta  ( v_m-i)  v_m-1}}\right)
\end{equation}
where $i=\sqrt{-1}$.
The leading-order behavior of the right hand side in the limit $\gamma,\,v_m\to\infty$ gives the asymptotic bound
\begin{equation}
Nu^{-1} \lesssim \frac{\pi}{k(\wRa Nu (1+2\Delta))^{1/2}}
\end{equation}

The resulting asymptotic lower bound on the Nusselt number is
\[Nu\gtrsim \frac{k^2(1+2\Delta)^2\wRa}{\pi^2}.\]
Clearly the bound increases with increasing $k$, and the lower bound can be increased by having $k$ scale with $\wRa$.
Taking $k\sim\wRa^\alpha$ results in the lower bound
\begin{equation}\label{eqn:NuBelow}
Nu \gtrsim \frac{(1+2\Delta)^2}{\pi^2}\wRa^{1+2\alpha}.
\end{equation}
However, it was shown in section \ref{sec:3A} that solutions do not exist for $\alpha<-1/2$ or $\alpha>1/4$, so the maximal lower bound is $Nu\gtrsim\wRa^{3/2}$, which occurs for wavenumbers $k\sim\wRa^{1/4}$ following the small-scale branch of the linear stability curve (\ref{eqn:LinStab}).

\subsection{Upper Bounds on $Nu$}
Upper bounds on the Nusselt number for these solutions can be obtained using the following upper bound to the radicand in the denominator of (\ref{eqn:NuV}), valid for large $v_m$ 
\begin{equation}
\ln\left(\frac{1+v_m^2}{1+v^2}\right)-\frac{k^4(v_m^2-v^2)}{\gamma}
\le (v_m^{-2}\ln(1+v_m^2)-k^4\gamma^{-1})(v_m^2-v^2).
\end{equation}
The resulting integral can again be evaluated in closed form, leading to
\begin{equation*}
Nu^{-1}\gtrsim \frac{\pi v_m}{k\gamma^{1/2}\left[(1+v_m^2)\left(\ln\left(v_m^2+1\right)-k^4v_m^2\gamma^{-1}\right)\right]^{1/2}}
\end{equation*}
Inserting the known bounds on $v_m$ (i.e.~equations (\ref{eqn:vmAbove}) and (\ref{eqn:vmBelow})) and using the fact that $v_m\to\infty$ to cancel factors in the numerator and denominator leads to
\begin{equation}
Nu^2 \lesssim \frac{k^2}{\pi^2}\left(\gamma\ln(k^{-4}\gamma)-\left(\frac{k^6\gamma}{k^6+8}-k^4\right)\right).
\end{equation}
Inserting the definition of $\gamma=\wRa Nu$ and using $k^4/Nu \le k^4$ yields
\begin{equation}
Nu \lesssim \frac{k^2}{\pi^2}\left(\wRa\ln(k^{-4}\wRa Nu)-\left(\frac{k^6\wRa}{k^6+8}-k^4\right)\right).
\end{equation}

Like the lower bound of the previous section, this upper bound depends on the scaling of $k$ with $\wRa$.
Allowing $k$ to vary as $k\sim\wRa^\alpha$ yields
\begin{equation}
Nu \lesssim \frac{1}{\pi^2}\left(\wRa^{1+2\alpha}\ln(\wRa^{1-4\alpha} Nu)-\left(\frac{\wRa^{1+6\alpha}}{\wRa^{6\alpha}+8}-\wRa^{4\alpha}\right)\right).
\end{equation}
The arguments of section \ref{sec:3A} show that there are no solutions for $\alpha<-1/2$ or $\alpha>1/4$, and the first term on the right hand side is clearly dominant for large $\wRa$ over this range of $\alpha$.
This leads to the bound
\begin{equation}
Nu \lesssim\frac{1}{\pi^2}\wRa^{1+2\alpha}\ln(\wRa^{1-4\alpha} Nu).
\end{equation}
These upper bounds add logarithmic corrections to the lower bound (\ref{eqn:NuBelow}).
The largest upper bound occurs for $k\sim\wRa^{1/4}$, where the dominant behavior is $Nu\lesssim \wRa^{3/2}\ln(Nu)$.
It should be noted that this `logarithmic correction' is not of the form $Nu\sim\wRa^{3/2}\ln(\wRa)$, but it is easy to verify that it implies $Nu\lesssim\wRa^{3/2+\eps}$ for any $\eps>0$.

\section{Numerical Solutions\label{sec:Num}}
\begin{figure*}
\begin{center}
\includegraphics[width=\textwidth]{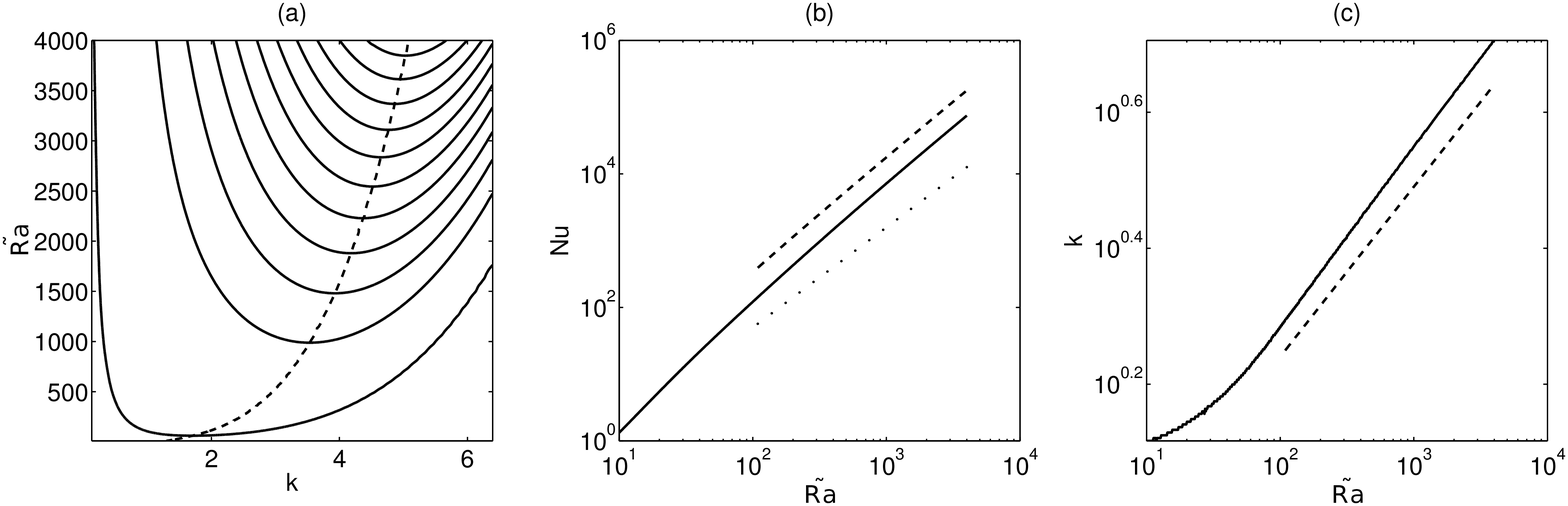}
\caption{(a) Contours of $Nu$ versus $k$ and $\wRa$; the contour interval is 7000 with a lowest contour of $100$; the dotted line shows the value of $k$ that maximizes $Nu$ at each $\wRa$. (b) Nusselt number as a function of $\wRa$ along the optimal value of $k$ from (a); the dashed line shows the behavior $\sim \wRa^{3/2}(\ln(\wRa))^{5/4}$ and the dotted line shows the lower bound $\sim\wRa^{3/2}$. (c) Optimal wavenumber $k$ as a function of $\wRa$; the dashed line shows the theoretical value $\sim \wRa^{-1/4}$. \label{fig}}
\end{center}
\end{figure*}
This section briefly presents some numerical solutions of the single mode equations (\ref{eqn:SM}) and (\ref{eqn:Nu}).
The focus is on the relationship between $\wRa$ and $Nu$; for the vertical structure of solutions see Refs.~\onlinecite{BZ94,SJKW06,JK07}.

Equation (\ref{eqn:SM}) is solved using Matlab's boundary-value solver bvp5c for specified $k$ and $\gamma$, and $\wRa$ and $Nu$ are then backed out from the solution using equation (\ref{eqn:Nu}).
Solutions are found for $50$ equally spaced wavenumbers from $k_c/10\approx 0.13$ up to $5k_c\approx 6.5$, and for Rayleigh numbers from critical up to $\wRa=4000$.
The solver requires an initial guess of the solution, to which the results are fairly sensitive.
The solution is initialized using $W=\sin(\pi z)$ at the smallest value of $k$ and a $\gamma$ 25\% above the local critical value, and is then continued to larger $k$ and $\gamma$.
At high $k$ and $\wRa$ the solution requires extremely high resolution, with the solver automatically generating up to $30,000$ points on the interval $z\in[0,1]$.
Although the solution does develop thin boundary layers and an isothermal interior (not shown), the majority of grid points chosen by the solver lie near the middle of the layer.
This is natural when viewed from the perspective of equation (\ref{eqn:ZInt}): the points are clustered near the singularity of the integral.

Figure \ref{fig}a shows a contour plot of $Nu$ over a range of $k$ from $k_c/10$ to $5 k_c$ and from $\wRa=10$ to $\wRa=4000$; the contour interval is $7000$.
The dashed line shows the value of $k$ that maximizes the Nusselt number at each $\wRa$.
The Nusselt number increases with $\wRa$, and is optimized by a value of $k$ that increases with $\wRa$.

Figure \ref{fig}b shows the maximum $Nu$ as a function of $\wRa$ in a log-log plot. 
The scaling $Nu\sim \wRa^{3/2}(\ln(\wRa))^{5/4}$ is shown by a dashed line, and the lower bound $Nu\gtrsim \wRa^{3/2}$ is shown by the dotted line.
Although the range of data is insufficient to draw precise conclusions, it appears that the Nusselt number grows slightly faster than $\wRa^{3/2}$.
Results at fixed $k$ show that the Nusselt number typically increases rapidly from the onset of convection and then settles down to a scaling somewhat closer to $Nu\sim\wRa\ln(\wRa)$ than to $Nu\sim\wRa$ (not shown).

Figure \ref{fig}c shows the value of $k$ that maximizes the Nusselt number as a function of $\wRa$ in a log-log plot.
The fastest-growing lower bound derived in the previous section was achieved for $\alpha\sim\wRa^{1/4}$, which is shown by the dashed line in Figure \ref{fig}c.
The agreement is quite close, although the range of data is again insufficient to draw precise conclusions.

These numerical results are broadly in agreement with the analysis of the previous section.

\section{Discussion}
In summary, upper and lower bounds on the Nusselt number associated with steady exact solutions of the non-hydrostatic quasigeostrophic equations\cite{JKW98,SJKW06} have been derived in the limit of large Rayleigh numbers.
The Nusselt number depends on the scaled Rayleigh number $\wRa=Ra E^{4/3}$ and on the wavenumber $k$ associated with the horizontal structure of the solutions.
For $k$ independent of $\wRa$ the lower and upper bounds are $\wRa \lesssim Nu\lesssim \wRa\ln(\wRa Nu)$ (constant prefactors are ignored in this section for clarity); the upper bounds at fixed $k$ have been derived previously.\cite{BZ94,JK99}
The bounds vary if the wavenumber $k$ is allowed to depend on the scaled Rayleigh number as $k=\wRa^\alpha$.
For large $k$ the upper and lower bounds are separated only by logarithmic factors.
The maximum possible lower bound is $Nu\gtrsim \wRa^{3/2}$ for $k\sim \wRa^{1/4}$, and the associated upper bound is $Nu \lesssim \wRa^{3/2}\ln(Nu)$, which is asymptotically smaller than $\wRa^{3/2+\eps}$ for any $\eps>0$.
Numerical solutions find that the Nusselt number tends to lie closer to the upper bounds than to the lower bounds for $\wRa$ up to $4000$, and that the optimal $k$ scales as $\wRa^{1/4}$.
This scaling of the wavenumber with Rayleigh number was also found to be optimal in numerical studies of the unreduced Boussinesq system using a variational upper-bound approach.\cite{Vitanov03,Vitanov10}

Rigorous upper bound theory for convection\cite{Howard63,Busse69,DC01} has difficulty with rotating Rayleigh-B\'enard convection because the methods typically rely on energy integrals, which are not affected by rotation.
Progress can be made using these methods at infinite Prandtl number since the velocities become slaved to the temperature through a linear operator that includes the effect of rotation.
These methods have not yet been applied to the NHQGE, but there are results for the unreduced equations.
In Ref.~\onlinecite{DC01} it was proven that $Nu\le c Ra^{2/5}$ for a constant $c$ independent of the rotation rate.
The alternative bound $Nu\le c(RaE^{-1}+2)^{4/11}$ for the unreduced system was also derived in Ref.~\onlinecite{Yan04}. 
The single mode solutions of the NHQGE are valid for any Prandtl number, including infinite, which suggests a conflict with the bounds quoted above.
However, some care must be taken in comparing these results to solutions of the NHQGE.

The NHQGE are derived as the leading-order behavior of an asymptotic expansion in powers of $E^{1/3}$; the prima facie assumption is thus that the scaled Rayleigh number $\wRa$ must be order-one with respect to $E^{1/3}$, i.e.~$\wRa\ll E^{-1/3}$, which by the definition of $\wRa$ implies $Ra\ll E^{-5/3}$.
There is evidence that the rotational constraint is lost at smaller $Ra$ though.
It is argued in Ref.~\onlinecite{JKRV12} that the breakdown occurs for $Ra\gtrsim E^{-8/5}$ and in Ref.~\onlinecite{KSB13} the breakdown is found to occur for $Ra\gtrsim E^{-3/2}$ on the basis of simulations of the unreduced equations. 

The bounds in Refs.~\onlinecite{DC01,Yan04} combined with the behavior of the single mode solutions effectively imply constraints on the range of $Ra$ and $E$ for which the single mode NHQGE solutions are permissible approximations of unreduced Boussinesq solutions.
The bound $Nu\lesssim Ra^{2/5}$ is only compatible with the behavior $Nu \sim \wRa^{3/2}=Ra^{3/2}E^2$ if $Ra<E^{-20/11}$.
This exponent of $\approx-1.82$ is consistent with both the prima facie estimate of $Ra<E^{-5/3}$ and the stricter, physically-motivated predictions of Refs.~\onlinecite{JKRV12,KSB13}.
The bound $Nu\lesssim Ra^{4/11}E^{-4/11}$ is even less restrictive since it is compatible with the behavior $Nu \sim \wRa^{3/2}$ for $Ra<E^{-52/25}$.
This exponent of $-2.08$ is well within the expected range of validity of the NHQGE.

Rigorous upper bounds for the infinite-Prandtl number NHQGE have recently been derived in Ref.~\onlinecite{GW14}.
The upper bound is of the form $Nu\lesssim\wRa^3$, which is consistent with the scaling conjectured in Refs.~\onlinecite{KSNUA09,KSA12}.
Simulations of the NHQGE display slower increase, on the order of $Nu\sim\wRa^{3/2}$, or at most $Nu\sim \wRa^2$ for infinite Prandtl number convection.\cite{SJKW06,JRGK12,JKRV12}
The solutions examined here correspond to laminar flow and are presumably more efficient (generate larger Nusselt numbers) than the turbulent solutions to which they are generally unstable.\cite{SJKW06}
It is possible that different laminar solutions might generate a larger heat flux; the convective Taylor columns of Ref.~\onlinecite{GJKW10} are a potential example.
But these columns have also been found\cite{JRGK12} to become unstable to turbulent dynamics at sufficiently large $\wRa$.
The behavior of the single mode solutions examined here suggests that the upper bound $Nu\lesssim\wRa^3$ from Ref.~\onlinecite{GW14} is pessimistic.

\begin{acknowledgments}
The author gratefully acknowledges improvements in presentation suggested by K.~Julien, and thanks G.~Vasil for pointing out a flaw in the original version of section \ref{sec:3A}.
\end{acknowledgments}

\bibliography{Draft_v2}
\end{document}